\documentstyle[sprocl]{article}
\def\Journal#1#2#3#4{{#1} {\bf #2}, #3 (#4)}

\def\nuc#1#2{$^{#1}{\rm #2}$}
\def\sles{\lower2pt\hbox{$\buildrel {\scriptstyle <}
   \over {\scriptstyle\sim}$}}
\def\sgreat{\lower2pt\hbox{$\buildrel {\scriptstyle >}
   \over {\scriptstyle\sim}$}}

\def\be{\begin{equation}}
\def\ee{\end{equation}}
\def\bea{\begin{eqnarray}}
\def\eea{\end{eqnarray}}

\begin{document}

\title{A Critique of Core--Collapse Supernova Theory Circa 1997}
\author{Adam Burrows}
\address{Department of Astronomy, University of Arizona, Tucson, AZ 85721, USA}

\maketitle\abstracts{There has been a new infusion of ideas in the study of the
mechanism and early character of core--collapse supernovae.  However, despite
recent conceptual and computational progress, fundamental questions 
remain.   In this all--too--brief contribution, I summarize some of the
interesting insights achieved over the last few years.  In the process, I highlight 
as--yet unsolved aspects of supernova theory that continue to make
it a fascinating and frustrating pursuit.}

\section{Introduction}

It has recently been shown that neutrino--driven Rayleigh--Taylor instabilities between
the stalled shock wave and the neutrinospheres (``Bethe'' convection) are generic feature of core--collapse
supernovae.\cite{bethe90,hbc92,hbhfc94,bhf95,jm96,mz96}  Whatever their role in reigniting the stalled explosion, their existence
and persistence have altered the way modelers approach their craft.  Supernovae must explode
{\it aspherically}, and this broken symmetry is stamped on the ejecta and character of the blast,
as well as on its signatures.   Consequences of asphericity include significant gravitational radiation,\cite{bh96,msmk91}
natal kicks to nascent neutron stars,\cite{bh96,w87} induced rotation,
\cite{bhf95} mixing of iron--peak and r--process nucleosynthetic products,
the generation and/or rearrangement of pulsar magnetic fields,
and, in extreme cases, jetting of the debris.  

However, there is no consensus yet on the centrality of overturn (or ``convection'') to the
mechanism of the explosion itself, with some deeming it either pivotal,\cite{bethe90,hbc92} 
potentially important,\cite{bhf95,jm96,mw88} or diversionary.\cite{mz96}  Nevertheless, {\it all} agree on the existence
of convection in the gain region of the stalled protoneutron star, and this point must be stressed.
{\em A gain region is a prerequisite for the neutrino--driven mechanism.\cite{bw85}  For heating to exceed 
cooling in steady--state accretion, the 
entropy gradient must be negative, and, hence, unstable.  Therefore, a gain region is always convective.}
In order to achieve quantitative agreement with the variety of observational constraints (explosion energy,
residual neutron star masses, $^{56,57}$Ni and ``$N = 50$'' peak yields, 
halo star element ratios,
neutron star proper motions, etc.), the ``final'' calculations 
must be done multi--dimensionally.
While if it can be shown that 1--D spherical models do explode after some delay, the true duration
of that delay, the amount of fallback, and the energetics of the subsequent explosion must be influenced by the overturning
motions that can not be captured in 1--D.  Convection changes not only the character of the hydrodynamics, but the entropies 
in the gain region and the ``efficiency'' \cite{hbc92} of neutrino
energy deposition that is the ultimate driver of the explosion.\cite{bhf95,janka93,bw85}  Furthermore, implicit in a focus
on 1--D calculations is the assumption that multi--D effects could only help, that they do not
thwart explosion.  
Hence, the belief that spherically--symmetric calculations are germane depends upon insights newly obtained
from the multi--dimensional simulations.  Nevertheless, it will be an important theoretical exercise to ascertain whether 
1--D models with the best physics and numerics can explode, if only because such has been a goal 
for decades.  The ``viablility'' of 1--D models will be influenced in part by the transport algorithm employed
(multi--group, flux--limited, full transport, diffusion), the microphysics (opacities and source terms
at high and low densities), the effects of general relativity, the equation of state, 
convection in the inner core that can boost the driving neutrino luminosities,\cite{b87,kjm96,mw88} and the inner density structure

\section{Neutrino Transport}

Though much of the recent excitement in supernova theory has concerned its
multi--dimensional aspects, neutrino heating and transport are still central
to the mechanism.  The coupling between matter and radiation in the semi--transparent
region between the stalled shock and the neutrinospheres determines the viability
and characteristics of the explosion.   Unfortunately, this is the most problematic regime.
Diffusion algorithms and/or flux--limiters do not adequately reproduce the effects of variations
in the Eddington factors and the spectrum as the neutrinos decouple.  Hence, a multi--group full transport
scheme is desirable.  

To address the issues surrounding neutrino transport, we have recently
created a neutrino transport code using the program {\bf Eddington} developed
by Eastman \& Pinto.\cite{ep93}  This code solves the full
transport equation using the Feautrier approach,
is multi--group, is good to order v/c, and does not employ
flux limiters.  The $\nu_e$s, $\bar{\nu}_{e}$s, and ``$\nu_\mu$s''
are handled separately
and coupling to matter is facilitated with accelerated lambda iteration (ALI).
By default, we employ 40 energy groups from 1 MeV to either 100 MeV ($\bar{\nu}_{e}$ and ``$\nu_\mu$s'')
or 230 MeV ($\nu_e$) and from a few to 200 angular groups, depending on the number of tangent rays at the given radial zone,
in the Feautrier manner.   In this way, the neutrino angular distribution function
and all the relevant angular moments (0'th through 3'rd) are calculated to high precision, for every energy group.

The effect of the full Feautrier
scheme vis--a--vis previous\cite{bhf95,mb93a,mb93b,mz96,w85,mb90} calculations
will soon be benchmarked and calibrated.
However, we have already obtained several interesting results.
Since the annihilation of $\nu$--$\bar{\nu}$ pairs into $e^+$--$e^-$ pairs depends upon the 0'th, 1'st, and 2'nd
angular moments of the neutrino angular distribution function, as well as upon the neutrino spectra, we can and have
calculated the rate of energy deposition via this process {\it exactly}, though in the context of previous model runs\,\cite{bhf95} 
(still ignoring general relativity).
The $\nu_e + \bar{\nu}_e \rightarrow  e^+ + e^-$ and $\nu_\mu + \bar{\nu}_\mu \rightarrow  e^+ + e^-$
energy deposition rates in the shocked region are no more than 0.01 and 0.001, respectively, those of the dominant charged--current
processes,  $\nu_e + n \rightarrow  e^- + p$ and $\bar{\nu}_e + p \rightarrow  e^+ + n$.  However, in the 
unshocked region ahead of the shock, depending upon the poorly--known $\nu$--nucleus
absorption rates, the $\nu$--$\bar{\nu}$ annihilation rate can be competitive, though
it is still irrelevant to the supernova.  These calculations should put to rest the notion 
that $\nu$--$\bar{\nu}$ annihilation
is important in igniting the supernova explosion.

It is thought that neutrino--electron scattering and inverse
pair annihilation are the processes most responsible for the energy
equilibration of the $\nu_\mu$'s and their emergent spectra.  However,
recent calculations imply that the inverse of nucleon--nucleon
bremsstrahlung ({\it e.g.}, $n + n \rightarrow n + n + \nu \bar{\nu}$) is
also important in equilibrating the $\nu_\mu$'s.\cite{suzuki}  This
process has not heretofore been incorporated in supernova simulations.
Our preliminary estimates suggest that inverse bremsstrahlung
softens the emergent $\nu_\mu$ spectrum, since the bremsstrahlung 
source spectrum is softer than that of pair annihilation.
In addition, given the large $\nu_\mu$ scattering albedo, one must 
properly distinguish absorption from scattering, in ways not possible 
with a flux--limiter.  Since 
the relevant inelastic neutral--current processes are stiff functions of neutrino
energy, these transport issues bear directly upon the viability of neutrino 
nucleosynthesis ({\it e.g.}, of \nuc{11}{B} and \nuc{19}{F}).\cite{whhh90} 

The new code allows us to calculate the difference bewteen the flux spectrum ($h_{\nu}$) and the
energy density spectrum ($j_{\nu}$).  The latter couples to matter and drives the supernova in the
neutrino mechanism,  while the
former, or some variant of it, is frequently substituted for the latter in diffusion
codes.  Since matter--neutrino cross sections are higher for higher--energy
neutrinos, the energy density spectrum is always harder than the flux spectrum.
This hardness boosts the neutrino heating rates in the semi--transparent
region.   To illustrate this effect, in Figure 1 the ratio $j_{\nu_e}$/$h_{\nu_e}$ is plotted versus neutrino energy at
a time 30 milliseconds after bounce.   The shock is then at 124 kilometers.  It is clear that
the ratio effect can be interesting.  However, it is most pronounced in the cooling region below
the gain region and tapers off as the shock is approached.  Mezzacappa {\em et al.},\cite{mz96}
in particular, have highlighted this correction, but self--consistent calculations from
collapse to explosion, using the Feautrier or Boltzmann techniques (in
principle equivalent), are needed, given the notorious feedbacks in the supernova problem.  
The same effect may be important in driving the protoneutron star wind\,\cite{bhf95} 
thought to be the site of the r--process.\cite{wh92,qian}
Indeed, full transport calculations of r--process winds and the supernova, even in 1--D, 
will be illuminating.

\begin{figure*}
\vspace*{2.5in}
\hbox to\hsize{\hfill\includegraphics{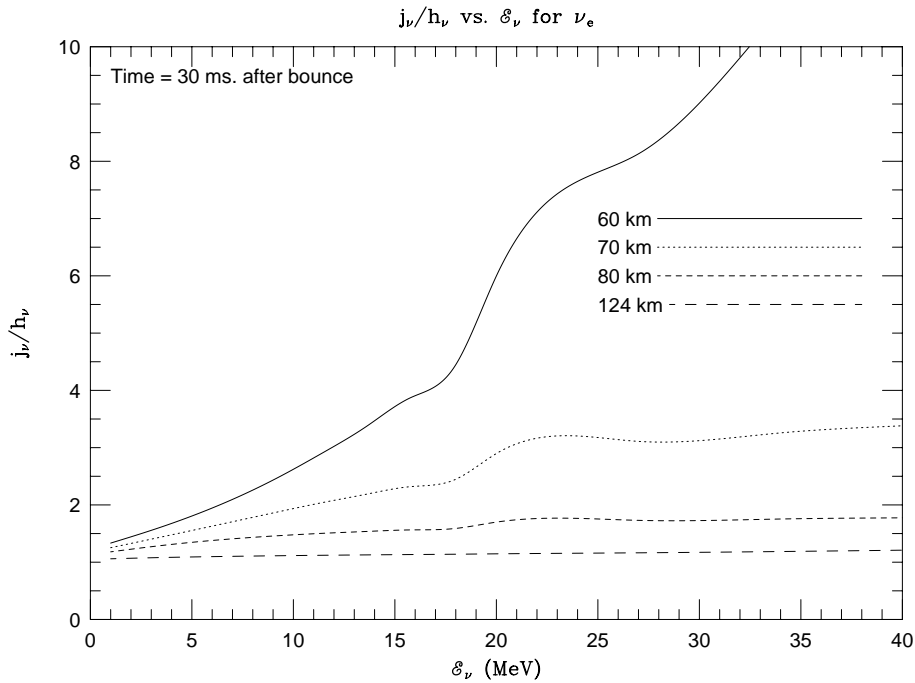}\kern+0in\hfill}

\caption{$j_{\nu}/h_{\nu}$ versus $\epsilon_\nu$ for electron neutrinos 30 milliseconds after the bounce
of a 15 M$_\odot$ core, using the code of Burrows \& Pinto (1997) }  

\end{figure*}


\section{Conclusions}

In parallel with the ongoing evolution of the numerical tools being
brought to bear on the supernova problem is the emerging realization that
the systematics of the supernova phenomenon with progenitor is inching
closer into view.  As we unravel the mechanism, we simultaneously explore
the origin of neutron stars and black holes, the birthplace of 
elements of which we are made, and the source of much of the energy of the ISM.
As supernova modelers and the Jayhawks might say, {\em ad astra per aspera}.

\section*{Acknowledgments} 
 
Conversations with Willy Benz, Chris Fryer, Tony Mezzacappa, and Phil Pinto that
materially altered the content of this squib are gratefully acknowledged, as is the support of the
U.S. N.S.F. under grant \#AST92-17322.

\end{document}